\documentclass[%
 amsmath,amssymb,
 aps,superscriptaddress,
pra,
floatfix,
reprint
]{revtex4-2}

\RequirePackage{graphicx}
\DeclareGraphicsExtensions{.pdf,.png,.jpg,.jpeg,.mps} 

\usepackage[
    type={CC},
    modifier={by},
    version={4.0},
]{doclicense}
\usepackage{orcidlink}
\usepackage{mhchem}
\usepackage{siunitx}
\RequirePackage{soul}
\RequirePackage[normalem]{ulem} 
\RequirePackage{hyperref} 
\definecolor{link_green}{rgb}{0.0,0.7,0.0}
\definecolor{link_blue_dark}{rgb}{0.0,0.0,0.7}
\definecolor{link_blue}{rgb}{0.0,0.0,1}
\definecolor{link_red}{rgb}{0.7,0.0,0.0}
\definecolor{link_red_dark}{rgb}{0.6,0.0,0.0}
\definecolor{link_red_verydark}{rgb}{0.3,0.0,0.0}
\hypersetup{hyperfigures=true,hyperfootnotes=true,hyperindex=true,pagebackref,breaklinks=true,citecolor=link_green,filecolor=link_blue,linkcolor=link_blue_dark, urlcolor=link_blue,colorlinks,bookmarks,bookmarksopen,bookmarksnumbered, pdfauthor="Olivier Fruchart",pdfstartview=FitH} \RequirePackage[all]{hypcap}

\RequirePackage{upgreek}%
\RequirePackage{journames}

\def\ie{{\it i.e.}\/\xspace}%
\def\etal{{\it et~al.}\/\xspace}%

\newcommand{\vect}[1]{\mathbf{#1}}
\newcommand{\unitvect}[1]{\vecthat{#1}}
\newcommand{\vecthat}[1]{\boldsymbol{\hat{\vect{#1}}}}
\newcommand{\unitvectvarphi}{\hat{\boldsymbol{\upvarphi}}}

  \definecolor{olivier_comment}{rgb}{0.6,0.6,0.9}
  \definecolor{olivier_add}{rgb}{0.8,0.2,0.2}
  \definecolor{olivier_remove}{rgb}{0.5,0.5,0.5}

\begin{document}
\doclicenseThis 

\title{Experimental determination and micromagnetic analysis of spin wave modes in cylindrical nanowires}
\author{Niklas Martin\,\orcidlink{0009-0008-5965-518X}}
\affiliation{Univ. Grenoble Alpes, CNRS, CEA, SPINTEC, Grenoble, France}
\affiliation{Univ. Grenoble Alpes, CNRS, Institut Néel, 38000 Grenoble, France}
\affiliation{Karlsruhe Institute of Technology, 76131 Karlsruhe, Germany}

\author{Laura Álvaro-Gómez\orcidlink{0000-0001-8899-3518}}
\affiliation{Dpto. de Física de Materiales, Universidad Complutense de Madrid, 28040 Madrid, Spain}

\author{Lucas Perez\orcidlink{0000-0001-9470-7987}}
\affiliation{IMDEA Nanociencia, Campus de Cantoblanco, 28049 Madrid, Spain}
\affiliation{Dpto. de Física de Materiales, Universidad Complutense de Madrid, 28040 Madrid, Spain}

\author{André Thiaville\orcidlink{0000-0003-0140-9740}}
\affiliation{Laboratoire de Physique des Solides, Université Paris-Saclay, CNRS, 91400 Orsay, France}

\author{Jean-Paul Adam\orcidlink{0000-0003-2025-7105}}
\affiliation{Université Paris-Saclay, CNRS, C2N, 91120 Palaiseau, France}

\author{Olivier Fruchart\,\orcidlink{0000-0001-7717-5229}}
\affiliation{Univ. Grenoble Alpes, CNRS, CEA, SPINTEC, Grenoble, France}

\author{Aurélien Masseb\oe uf\orcidlink{0000-0003-4239-1313}}
\email{aurelien.masseboeuf@cea.fr}
\affiliation{Univ. Grenoble Alpes, CNRS, CEA, SPINTEC, Grenoble, France}

\date{\today}


\begin{abstract}
We report an experimental study of spin wave modes in individual cylindrical nanowires, a textbook situation of confined spin waves in 3D nanomagnetism. We observe discrete modes of thermal spin waves with micro-Brillouin light scattering, whose frequencies $f$ shift to higher values as the applied longitudinal induction magnetic field $B_z$ increases. Micromagnetic simulations allowed us to associate every $f(B_z)$ curve to a given spatial mode, labeled with radial and azimuthal indices $\ell$ and $m$.
\end{abstract}


\maketitle

\let\clearpage\relax

\section{Introduction}
\label{sec:introduction}

Magnonics considers magnetization dynamics in the form of spin waves. These collective excitations of magnetic moments give rise to a rich physics, such as propagating and standing spin waves, hybridization between modes, similarity with Bose-Einstein condensation and coupling with the lattice~\cite{bib-STA2009,bib-GUR2020}. They also provide the basis for magnonic logic gates~\cite{lenk_building_2011,sebastian_micro-focused_2015} and non-conventional logic such as reservoir computing relying on spin waves~\cite{bib-NAK2018,bib-MAH2020,bib-GAR2022}. Most of these concepts for devices are based on the propagation of spin waves in planar magnetic wave guides, especially quasi-one-dimensional (1D) geometries like micro- or nanostrips~\cite{demidov_nonlinear_2009}. At the same time, three-dimensional (3D) nanomagnetism is an emerging field coming with new physical effects related to the 3D shape of objects, new topologies and curvature~\cite{fernandez-pacheco_three-dimensional_2017}, the latter liable to give rise to non-reciprocal spin-wave dispersion curves~\cite{otalora_curvature-induced_2016,otalora_asymmetric_2017}. It is hoped that it can also provide opportunities for applications. It is in this context that magnonics in 3D is being developed~\cite{bib-GUB2019}.

Cylindrical nanowires are textbook 3D magnetic conduits: they can be seen as quasi-1D owing to their large aspect ratio, but yet display 3D characteristics if their diameter exceeds about ten times the dipolar exchange length~\cite{jamet_head-to-head_2015}. The rotational invariance of their cross section promises the physics to boil down to its simplest form, in contrast to more complex geometries such as with a hexagonal cross-section~\cite{korber_symmetry_2021}.
The theory of spin waves in an infinite cylindrical nanowire with axial magnetization at rest has been elaborated by successive works and is now rather extensive~\cite{bib-JOS1961,arias_theory_2001,kraus_ferromagnetic_2011,rychly_spin_2019}, providing analytical expressions for the radial and azimuthal components of the dynamic magnetization, as well as numerically computed dispersion relations for the axial propagation wave vector. In a seminal work, Joseph \etal considered magnetostatic modes only, thus applicable to large wires~\cite{bib-JOS1961}. This gives rise to a negative group velocity, as the geometry is similar to backward-propagation volume modes~\cite{bib-STA2009,bib-GUR2020}. Arias and Mills considered both exchange and magnetostatics, suitable to describe nanowires~\cite{arias_theory_2001}, predicting a negative group velocity for low $k$ and a positive one for larger~$k$. More recent works provide numerous views and details about radial and azimuthal modes and their hybridization~\cite{rychly_spin_2019}. The interest in spin waves in nanowires has been revived recently, owing to the prediction of their emission by solitons, such as domain walls, upon motion at speed above typically \qtyrange{500}{1000}{\meter\per\second}, playing a crucial role in the limitation of their speed~\cite{thiaville_domain-wall_2006,yan_fast_2011,yan_spin-cherenkov_2013}. Besides fundamental interest in the coupling of a moving soliton with spin waves, this comes in the applicative context of the proposal to use domain walls as means to store information in two- or three-dimensional race-track memories~\cite{parkin_magnetic_2008}.

In contrast, experimental reports are rather incomplete at this stage. Ebels \etal conducted ferromagnetic resonance~(FMR) experiments on assemblies of Ni nanowires in low-density polycarbonate matrices, with diameter in the range \qtyrange{35}{500}{\nano\meter}~\cite{bib-EBE2001b}. Spectra are consistent with the uniform Kittel mode, however sub-structure peaks were reported. These had not been formally interpreted but it was mentioned that they may be related to the exchange-dipolar spin wave modes predicted not long before~\cite{arias_theory_2001}. Wang \etal reported Brillouin Light Scattering of thermal spin wave modes on assemblies of Ni nanowires in large-density alumina matrices, with diameter in the range \qtyrange{25}{55}{\nano\meter}~\cite{wang_spin-wave_2002}. Multiple peaks were also reported, consistent with exchange-dipole spin wave mode with a radial profile, as their splitting increased with decreasing diameter. However, the sizable strength of inter-wire interactions was acknowledged, consistent with existing knowledge~\cite{encinas-oropesa_effect_2001}. These made the identification of the modes challenging, and an exchange stiffness much smaller  up to one order of magnitude below the bulk value was required to account for the peak positions. The many possible physical grounds for multi-peak FMR spectra in nanowire arrays were discussed in Ref.~\citenum{bib-KRA2017}, in the context of experimental spectra of arrays of \qty{40}{\nano\meter} Co nanowires. A similar conclusion was reached that the understanding of the several peaks on the basis of spin wave modes would imply a value of exchange stiffness much lower than expected, and that the role of distributions and inter-wire interaction was prominent. In all these cases the distribution of internal and interaction properties within the large set of wires probed is liable to contribute to the peak broadening and it is challenging to ascribe each of them to a given mode. Investigation of spin waves in individual cylindrical structures is emerging, with pioneering reports of strip-line-excited spin waves in hexagonal nanotubes~\cite{korber_symmetry_2021,giordano_confined_2023}. Indications of non-reciprocity were given, associated with curvature, however the lack of rotational symmetry for both the hexagonal tube and the excitation source led to a rich spectrum of spin waves, but difficult to interpret fully.

In this manuscript we report the investigation of thermal spin waves in individual \ce{Fe_{20}Ni_{80}} nanowires using micro-focused  Brillouin light scattering~(µBLS). We could ascribe each of the seven distinct peaks measured to a given spin wave mode, labeled with azimuthal and radial indices. The good fit of all modes over a large range of applied longitudinal field, and resulting in a realistic set of material parameters, raises confidence about their unambiguous identification. The manuscript is structured as follows. In sec.~\ref{sec:sample-fabrication} we describe the fabrication process of cylindrical ferromagnetic nanowires and the  preparation for single  wire measurements as well as the techniques implemented. The experimental results, the peak identification with simulations and discussion are presented in sec.~\ref{sec:experimental-results}.

\let\clearpage\relax

\section{Methods}
\label{sec:sample-fabrication}

\begin{figure}
	{\centering
	\includegraphics[width=8.6cm]{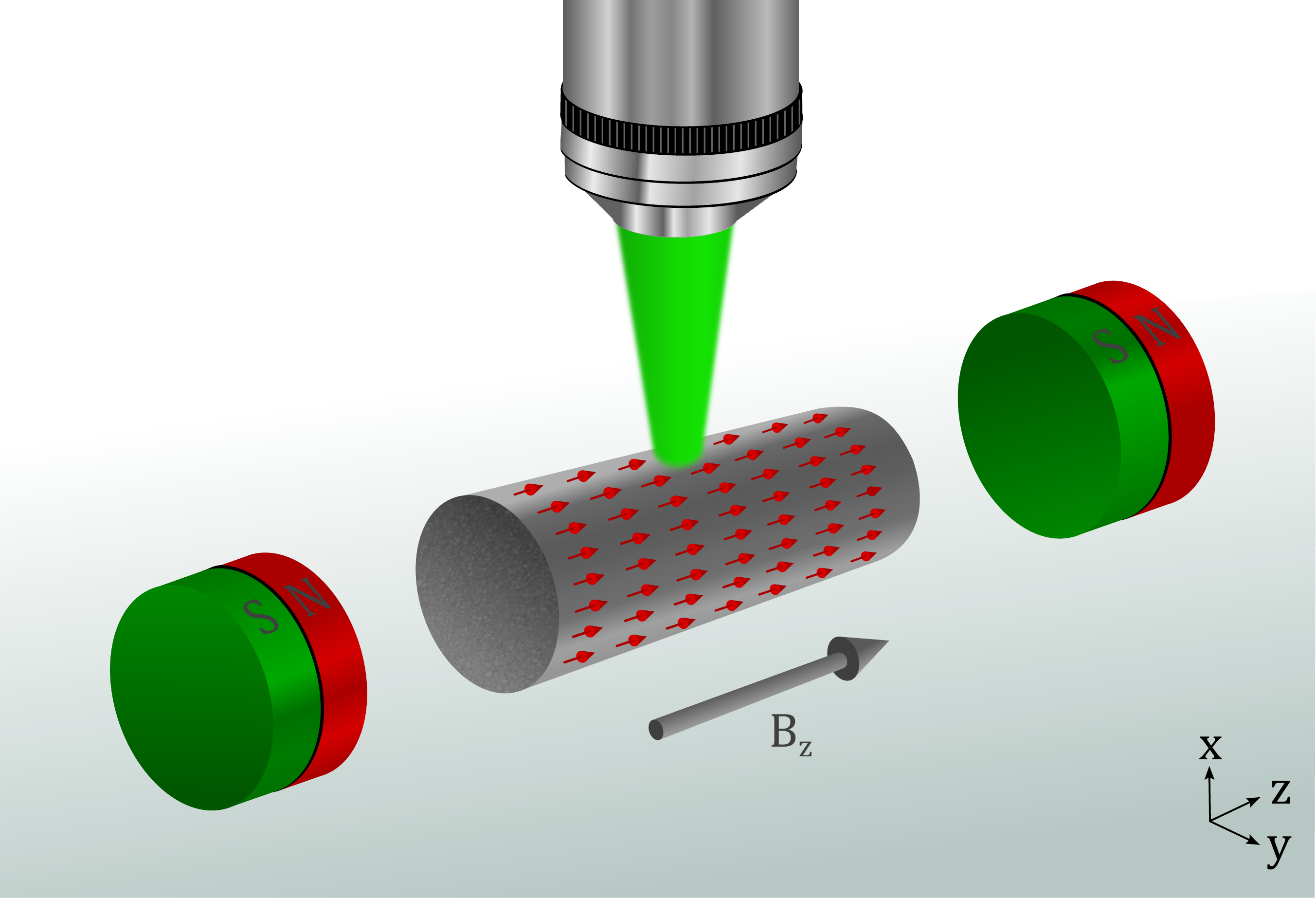}}
	\caption{\label{fig:bls-thermal-setup}Sketch of the µBLS setup (not to scale). A nanowire with axial magnetization (red arrows) is exposed to a Laser beam (green) focused by a microscope objective. A set of two permanent magnets allows to apply a longitudinal magnetic field to the sample. The distance between the magnets and the sample sets the magnitude of the applied magnetic field.}
\end{figure}

We used nanowires made of permalloy (Py), with targeted composition \ce{Fe_{20}Ni_{80}}. These are grown by electrochemical deposition in porous alumina templates, with typical length of several tens of micrometers. In this work, we consider wires with diameter $\SI{115}{nm}$, and measured composition \ce{Ni_80 Fe_20}. More details can be found elsewhere~\cite{bib-RUI2018}. In order to investigate a single nanowire, the alumina template is dissolved in a \ce{H3 PO4} ($\SI{0.4}{M}$) \ce{H2 CrO4} ($\SI{0.2}{M}$) solution, then rinsed several times in acetone and ethanol for purification. The solution containing the freed nanowires is then spread onto either a Si substrate.

Brillouin Light Scattering enables one to investigate magnetic excitations either of thermal nature or excited on purpose~\cite{sebastian_micro-focused_2015}. It is based on inelastic scattering of light, on magnons, the quasiparticles of spin waves. The scattering process can either create or annihilate a magnon, thereby either decreasing or increasing~(resp.) the energy of the photon by the energy of a magnon. These two processes are commonly called Stokes and anti-Stokes, respectively. The combination of spectroscopy and angular variation allows one to derive the spin wave dispersion curve $f(k)$, based on the conservation of both energy and momentum. The illumination is generally from a laser with a typical spot size of a few tens of micrometers. The resulting low filling factor leads to spectra with peaks barely discernable from the background (not shown). In micro-focused BLS~(µBLS) the laser beam is focused to a sub-micrometer spot size by a microscope objective. In contrast to regular BLS, µBLS has therefore a higher spatial resolution, but at the expense of wave vector resolution, prevented by the large numerical aperture of the objective and the Heisenberg's uncertainty principle~\cite{sebastian_micro-focused_2015}. The µBLS setup used in this work has a laser wavelength of $\SI{532}{nm}$, power set to $\SI{1.5}{mW}$, and incidence normal to the supporting surface, as sketched in Fig.~\ref{fig:bls-thermal-setup}. This collects wave vectors in the maximum range $\qty{\pm17.8}{\radian\per\micro\meter}$. A Tandem Fabry-Pérot interferometer (The Table Stable ltd.) measures the frequency of the inelastically back-scattered light. Permanent magnets mounted on translating stages on either side of the sample allow to apply a variable magnetic induction field~$B_z$ up to $\qty{0.2}{\tesla}$. We considered a wire with diameter $\SI{115}{nm}$ with axis aligned with the applied magnetic field. Spectra are acquired for applied induction field up to $\qty{200}{mT}$, in steps of $\SI{25}{mT}$. Each spectrum results from $\SI{24}{h}$ acquisition time. To extract the frequency of every mode, we fit the average  of the Stokes and anti-Stokes parts with a weighted sum of Lorentzian functions, naturally suitable to describe the dynamic susceptibility.

Micromagnetic simulations were performed using \texttt{TetraX}, a finite-element dynamical matrix-based software package~\cite{korber_finite-element_2021,tetrax,korber2021numerical}. \texttt{TetraX} allows one to compute spin waves dispersion curves in systems geometrically invariant upon translation along a given direction, typically nanowires and nanotubes. The calculation takes place in a cross-sectional 2D space, but can describe spin waves with an axial wave vector~$k_z$. It enables one to compute dispersion relations $f(k_z)$ and mode profiles in the cross-section, with $z$ the axial coordinate, as well as their magnetic-field dependence. We considered a diameter $d=\SI{115}{nm}$, discretized in a finite-element mesh of size $\SI{4}{nm}$. We used a damping parameter of $\alpha=0.01$ and the gyromagnetic ratio is set to $\gamma/2\pi=\SI{28.02}{GHz/T}$. The latter has been used by some to simulate the Permalloy material~\cite{sandweg_modification_2008}. However, the Landé factor of Permalloy has been measured precisely to be $g=2.11$~\cite{bib-SHA2013b}, which translates to $\gamma/2\pi=\SI{29.6}{GHz/T}$. Considering the latter would slightly change the material parameter resulting from the fitting procedure described in the results section.

\let\clearpage\relax

\section{Results}
\label{sec:experimental-results}

\begin{figure}[t]
    {\centering
    \includegraphics[width=8.6cm]{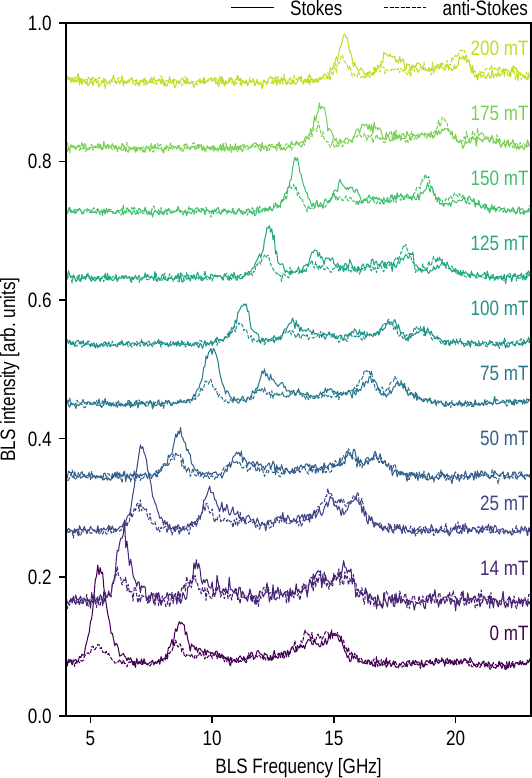}}
    \caption{\label{fig:bls-spectra}Measured µBLS spectra of a single isolated nanowire for different $B_z$. The Stokes signal is folded onto the anti-Stokes signal.}
\end{figure}

\begin{figure}
	{\centering
	\includegraphics[width=8.6cm]{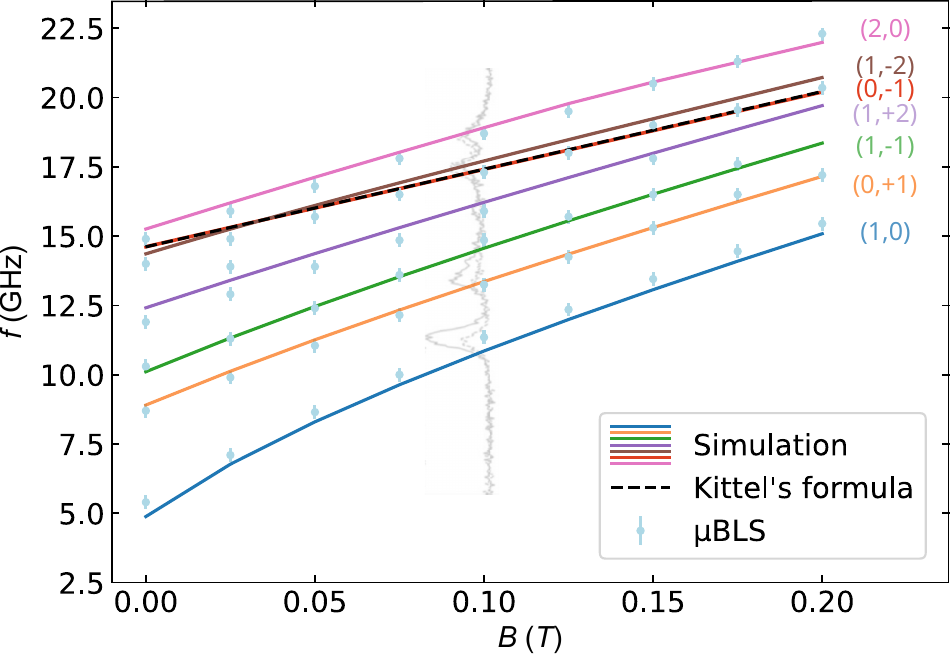}}
	\caption{\label{fig:simulation-exp-fB}Magnetic field dependence of spin wave modes. The points correspond to the µBLS spectra peaks obtain in Fig.~\ref{fig:bls-spectra} (an overlay of the spectra obtained at 75mT is displayed as guide to the eye). The error bars of extent $\SI{0.25}{GHz}$ reflect the limited spectral resolution of the setup, plus uncertainties of peak fitting~(see text). Simulated $f_{k=0}(B_z)$-curves for the seven first modes are displayed as colored solid lines (using same color code as in Fig.~\ref{fig:full-dispersion}). Kittel's analytical curve for the uniform mode is traced as a dotted black line and coincides perfectly with the simulation of mode $(0,+1)$.
	 }
\end{figure}

Fig.~\ref{fig:bls-spectra} shows the Stokes and anti-Stokes parts of µBLS spectra of a uniformly-magnetized $\qty{115}{nm}$-diameter \ce{Ni_80 Fe_20} nanowire, measured for a series of values of longitudinal external magnetic field up to $\qty{200}{mT}$. Each spectrum  in the series shows distinct peaks, whose frequencies depends on the magnitude of $B_z$. This shows the magnetic nature of the observed peaks, which we attribute to thermal spin wave modes. Anti-stokes peaks are sizable, which is expected as at room temperature the thermal energy $k_\mathrm{B}T\approx \qty{25}{meV}$ is much larger than the magnon energies $\approx \qty{0.01}{meV}$. Note that in conventional BLS an intensity difference between Stokes and anti-Stokes is nevertheless often seen, due to optical and magnetic effects~\cite{bib-ROU1995a}. The frequency of the modes increases with the applied field, which is expected as the magnetic field is applied along magnetization and the wire axis, adding up to the shape anisotropy field. The frequencies of the peaks versus applied field are displayed in Fig.~\ref{fig:simulation-exp-fB} as data points. Part of the error bars $\delta f$ relates to the spatial resolution of the Fabry-Pérot interferometer. Uncertainties also arise from the data fitting. For the main four peaks in Fig.~\ref{fig:bls-spectra}, \ie, with the largest intensity, the additional uncertainty ranges from \qtyrange{50}{150}{\mega\hertz}. The uncertainty for the less-intense peaks is larger, estimated to $\delta f=\qty{25}{\mega\hertz}$, which makes the error bars.

\begin{figure}
    {\centering
    \includegraphics[width=8.6cm]{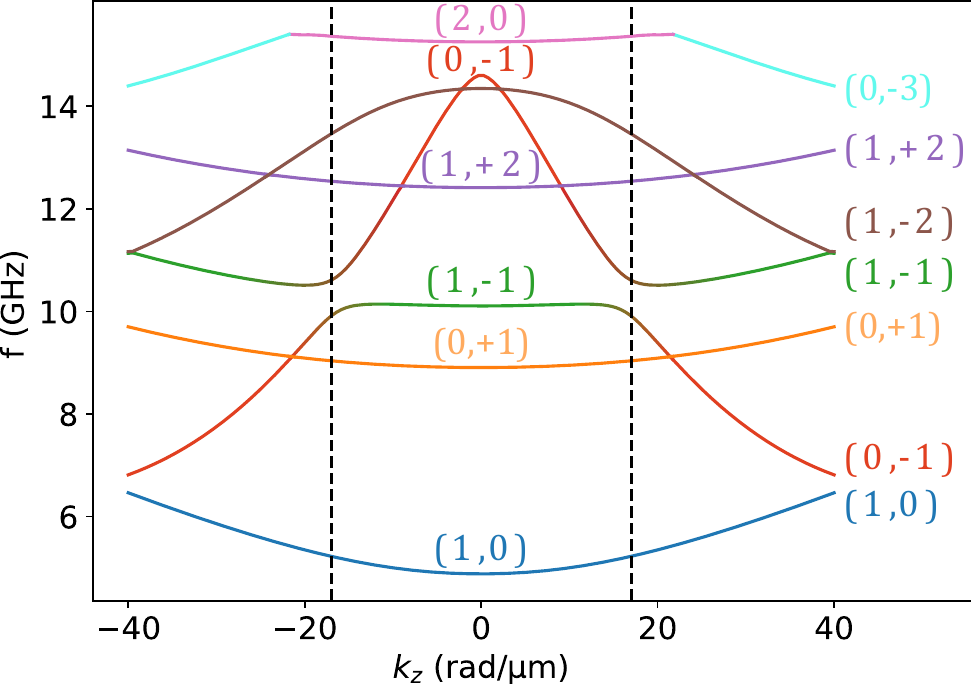}}
    \caption{\label{fig:full-dispersion}Simulated full spin wave dispersion $f_n (k_z)$ of a cylindrical nanowire with a diameter of $\SI{115}{nm}$ at $B_z=\SI{0}{mT}$, for $A_\mathrm{ex}=1.15 \times 10^{-11} \si{J/m}$ and $M_\mathrm{s}=\SI{830}{kA/m}$. The vertical dashed lines indicate the collection range of the µBLS objective. The labels of the modes $(i,j)$ are explained in the text.}
\end{figure}

We now turn to simulations to elucidate the nature of the spin-wave modes. Fig.~\ref{fig:full-dispersion} shows the computed spin wave dispersion curves $f(k_z)$ for the seven modes of lowest energy. Depending on the mode the group velocity $v_\mathrm{g}=\partial f/\partial k_z$ may be either positive or negative, depending whether they are exchange- or dipolar-dominated, respectively~\cite{arias_theory_2001}. Crossings and anti-crossings are expected when two modes have an identical energy, depending on whether the symmetry of the two modes allows mutual coupling~\cite{rychly_spin_2019}. Colors are ascribed not to an energy ranking but to a given physical mode, whose identification we detail below.


\texttt{TetraX} delivers the eigenvectors of the linearized
LLG equation. These describe the oscillatory part of magnetic moments, which can be illustrated within a single cross section of the nanowire at a given time. The oscillatory part is described by a real and an imaginary part~\cite{korber_finite-element_2021}, which can also be converted to the magnitude $m$ and the relative phase $\phi$ of the precession. A natural choice for decomposing the dynamic magnetization component is using the local polar coordinates $r$ and $\varphi$~(see the inset at the bottom-right of Fig.~5
):

\begin{equation}
        \phi_i = \arctan \left[\frac{\mathfrak{Im}(m_i)}{\mathfrak{Re}(m_i)}\right], \qquad i=r,\varphi
\end{equation}
\begin{equation}
        |m_i|=\sqrt{\mathfrak{Re}(m_i)^2 + \mathfrak{Im}(m_i)^2}, \qquad i=r,\varphi
\end{equation}
\begin{figure*}[!t]
    \centering
    \includegraphics[width=\textwidth]{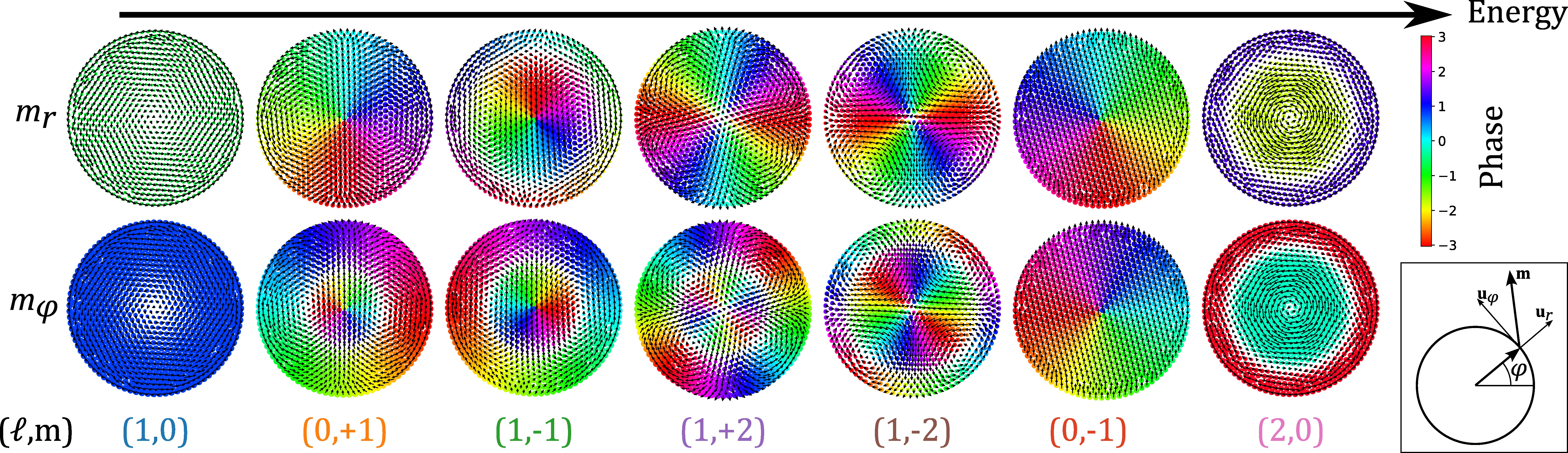}
    \caption{\label{fig:mr-mphi-amp-phase-7modes}Spatial mode profiles for the radial and azimuthal components of the dynamic magnetization for the first seven modes~(the bottom-right insert sketches the definition of the radial and azimuthal components of dynamic magnetization). For each component the phase is encoded as a colormap and the amplitude as the intensity of these colors, via the dot size.  The label $(\ell,m)$ corresponds to the number of radial and azimuthal nodes, see text for their definition, with colors matching the curves in Fig.~\ref{fig:full-dispersion}. Except for modes with $m=0$, which display rotational symmetry, the reference phase is set to zero for the radial component on the top periphery. These mode profiles are obtained using $A_\mathrm{ex}=1.15 \times 10^{-11} \si{J/m}$ and $M_\mathrm{s}=\SI{830}{kA/m}$.}
\end{figure*}
Fig.~5
shows the spatial profiles of the seven modes of lowest energy, \ie, describing the dynamic part of magnetization. For both their display and their labeling, we follow earlier works of precessional modes with rotational geometrical and magnetic invariance, either in perpendicularly-magnetized disks or in axially-magnetized infinite cylinders~\cite{naletov_identification_2011,rychly_spin_2019}. As regards the display, each row describes a given component of the oscillatory part of magnetization, the radial one $m_r$ in the first row and the azimuthal one $m_\varphi$ in the second row. For each of them, the phase is displayed with a color scheme. A positive value of phase indicates a delay, with the choice that the static part of magnetization $m_z$ points towards the reader to be consistent with the physical counterclockwise rotation. The amplitude of the component is displayed via the size of the colored dots such as done in Ref.~\citenum{giordano_confined_2023}. When averaged over several sites, the latter provides a visual perspective very similar to the intensity of the color~\cite{rychly_spin_2019,naletov_identification_2011}. For every mode, black arrows depict the real part of the vector magnetic moments standing for the distribution of the precessional moment at an arbitrary time $t=0$. This overlay is the same for both components, with a view to help identifying the modes. Let us take two examples. First, the mode in the first column has a rotational invariance, with signature the uniform color for each component. The $\pi/2$ phase shift between $m_r$ and $m_\varphi$ and the smaller amplitude of the former (see the color intensity) indicates an elliptical trajectory elongated parallel to the wire surface, \ie, along the azimuthal direction. This is similar to the Kittel FMR mode of thin films with in-plane magnetization, limiting the formation of surface charges. Second, the mode in column~6 is uniform, and precessional is circular as shown by the identical size of the colored dots. This is the uniform FMR mode.

Due to the potential crossing of energy of the modes versus $k$, diameter and applied magnetic field, it is helpful to follow and label each mode unambiguously. In the cylindrical geometry it is common to label the modes with the radial index $\ell=0,1,2,...$, reflecting the number of nodes of the radial amplitude along the radius, and index $m=0 \pm 1, \pm 2, ...$, reflecting the number of turns of the azimuthal dynamical magnetization component $m_\varphi$ around the periphery, in the local frame of the cylindrical leading vector $\unitvect r$ and $\unitvectvarphi$. This nomenclature is the basis of the mathematical functions used for analytical modeling, with separation of the radial and azimuthal variables~\cite{rychly_spin_2019}. For instance, the rotationally-invariant mode in column is labeled $(1,0)$, while the Kittel mode in column 6 is labeled $(0,-1)$. Besides Fig.~5
, in  appendix we provide a view of more modes for $\ell =0,1,2, 3$ and $m = \pm 0,1,2,3,4$~(Fig.~7).
An equivalent way to label these modes is to consider the winding number $w=m+1$, being indeed zero for the Kittel mode. Similar nomenclatures have been used for whispering gallery modes in disks~\cite{schultheiss_excitation_2019} and spin waves in nanotubes~\cite{korber_curvilinear_2022}, however for $m_z$ profiles. Note that other works define angles with respect to a fixed frame, in which case the zeroth azimuthal order stands for the Kittel mode~\cite{naletov_identification_2011,bib-TAU2016}.

With a view to analyze the experiments, we extended the simulations of the seven modes of lowest energy, considering also an axial external magnetic field $B_z$ in the range $\qtyrange{0}{200}{\milli\tesla}$. We sought to reproduce all seven experimental curves with a single set of parameters. We used $f(k_z=0)$ to compare with the experiments, expecting a peak in the density of states related to the zero slope of $f(k_z)$ at the origin. Note that µBLS peak shape modeling is under development, however is suitable only in the case of flat and extended multilayers~\cite{bib-WOJ2024}, where the additional complexities of optics and magneto-optics of curved samples are absent~\cite{bib-RIC2017}. Upon doing so, all simulated curves happen to depend on the system parameters~(wire diameter, magnetization and exchange stiffness), however each with a different impact depending how much exchange and magnetostatics are involved in the mode. However, it turns out that the diameter and exchange stiffness play rather correlated roles, and a broad range of values may allow a reasonable fit. Consequently we decided to keep the diameter set to $\qty{115}{\nano\meter}$, which we could measure accurately with scanning electron microscopy on the very same wire measured by µBLS, and let exchange be fitted, as its value is not known with high accuracy in magnetic materials. The best agreement was sought by iterative testing, and found for $M_\mathrm{s}=\qty{830}{kA/m}$ and $A_\mathrm{ex}=\qty{1.15e-11}{\joule\per\meter}$. The former is in excellent agreement with magnetization of bulk Permalloy. The latter value is in the range expected for Permalloy. Its determination via the present fitting is quite trustworthy as curves for non-uniform modes are very sensitive to $A_\mathrm{ex}$. The resulting curves $f(k_z=0,B_z)$ are plotted in Fig.~\ref{fig:simulation-exp-fB} in addition to the experimental data, for the seven modes of lowest energy modes displayed in Fig.~\ref{fig:mr-mphi-amp-phase-7modes}. Within the resolution of the µBLS setup, all spin wave modes predicted by the simulation are observed experimentally. As an additional check, we plotted on Fig.~\ref{fig:simulation-exp-fB} the Kittel mode as a dashed black line, which perfectly fits the simulated $(0, -1)$ mode:

\begin{equation}
    f=\frac{\mu_0 |\gamma|}{2\pi}\left(\frac{B}{\mu_0}+\frac{M_\mathrm{sat}}{2}\right)\;.
    \label{eq:kittel}
\end{equation}

\let\clearpage\relax

\section{Conclusion}
\label{sec:conclusion}

We investigated experimentally thermal spin waves in cylindrical nanowires with micro-BLS up to $\qty{23}{\giga\hertz}$, evidencing seven modes. Adjusting the variation of their frequency versus longitudinal applied field with simulations of eigenmodes using a single set of material parameters allowed us to ascribe each of them unambiguously to one of the seven eigenmodes of lowest energy, labeled with radial $\ell$ and azimuthal $m$ indexes. The fitting procedure provides an accurate value of exchange stiffness, similar to the case of standing spin waves in extended thin films. This demonstrates the ability to measure precisely spin waves locally in cylindrical wires, which opens the route to track the strong spin-wave emission predicted in these systems with large domain-wall mobility.

\section{Acknowledgements}

We acknowledge support from the Regional Government of Madrid under Project TEC-2024/TEC-380 (Mag4TIC-CM). We thank Daria \textsc{Gusakova} Nicolas \textsc{Jaouen}, Olivier \textsc{Klein}, Lucía \textsc{Gómez Cruz}, Christophe \textsc{Thirion} and Rachid \textsc{Belkhou} for helpful discussions, and the latter also for a critical reading of the manuscript. A CC-BY public copyright license has been applied by the authors to the present document and will be applied to all subsequent versions up to the Author Accepted Manuscript arising from this submission, in accordance with the grant’s open access conditions~\cite{bib-CC-BY}.

\let\clearpage\relax

\section*{Appendix}

In this appendix we show the energy~(Fig.~6 and Tab.~I) and spatial profile~(Fig.~7) of more modes, compared with those discussed in the manuscript to fit experimental data.

\begin{figure}[!h]
    {\centering
    \includegraphics[width=8.6cm]{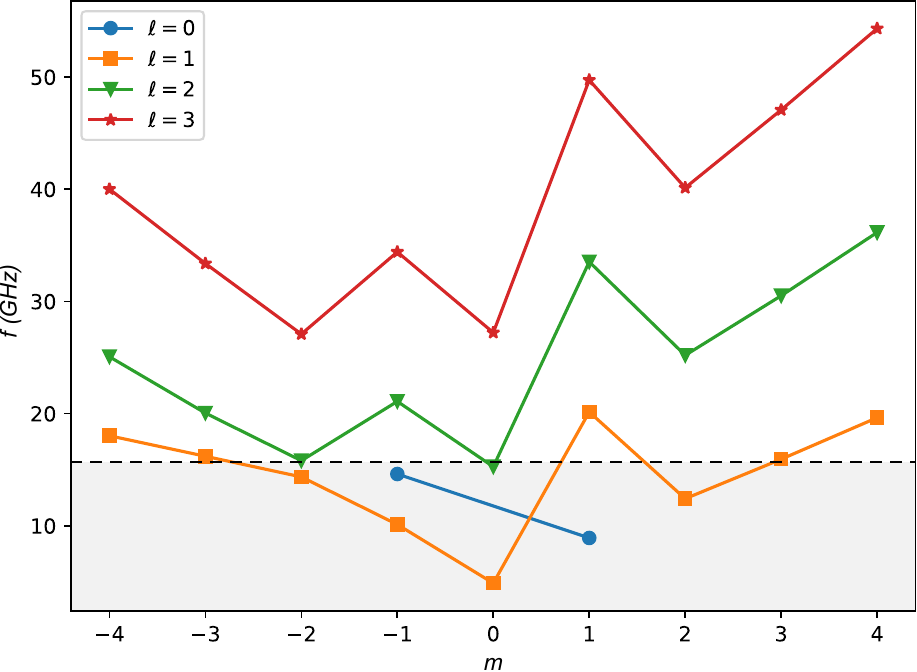}}
    \caption{\label{fig:all-modes-frequencies}Energy of the modes shown in Fig.~7
, sorted by their number of radial nodes $\ell$ and azimuthal nodes $m$. Their values are listed in Tab.~I
. The shaded area highlights the energy range of the modes shown in Fig.~5
.}
\end{figure}

\begin{table}[!h]
\centering
\caption{\label{tab-all-modes-frequencies}Energy of the modes shown in Fig.~7
\ and plotted in Fig.~6
.}
\begin{tabular}{cc}
\hline\hline
$(\ell, m)$ 	& 	$f$ (GHz) \\
\hline
$(0,-1)$ 	& 	14.61 \\
$(0,+1)$	& 	8.91 \\
$(1, -4 )$	& 	 18.02  \\
$(1, -3 )$	& 	 16.19  \\
$(1, -2 )$	& 	 14.35  \\
$(1, -1 )$	& 	 10.11  \\
$(1, 0 )$	& 	 4.88  \\
$(1, 1 )$	& 	 20.14  \\
$(1, 2 )$	& 	 12.41  \\
$(1, 3 )$	& 	 15.93  \\
$(1, 4 )$	& 	 19.63  \\
$(2, -4 )$	& 	 25.06  \\
$(2, -3 )$	& 	 20.04  \\
$(2, -2 )$	& 	 15.8  \\
$(2, -1 )$	& 	 21.08  \\
$(2, 0 )$	& 	 15.26  \\
$(2, 1 )$	& 	 33.52  \\
$(2, 2 )$	& 	 25.19  \\
$(2, 3 )$	& 	 30.5  \\
$(2, 4 )$	& 	 36.14  \\
$(3, -4 )$	& 	 40.0  \\
$(3, -3 )$	& 	 33.37  \\
$(3, -2 )$	& 	 27.08  \\
$(3, -1 )$	& 	 34.4  \\
$(3, 0 )$	& 	 27.21  \\
$(3, 1 )$	& 	 49.72  \\
$(3, 2 )$	& 	 40.12  \\
$(3, 3 )$	& 	 47.06  \\
$(3, 4 )$	& 	 54.31\\
\hline\hline
\end{tabular}
\end{table}

\newpage

\begin{figure*}[b!]
    {\centering
    \includegraphics[width=\textwidth]{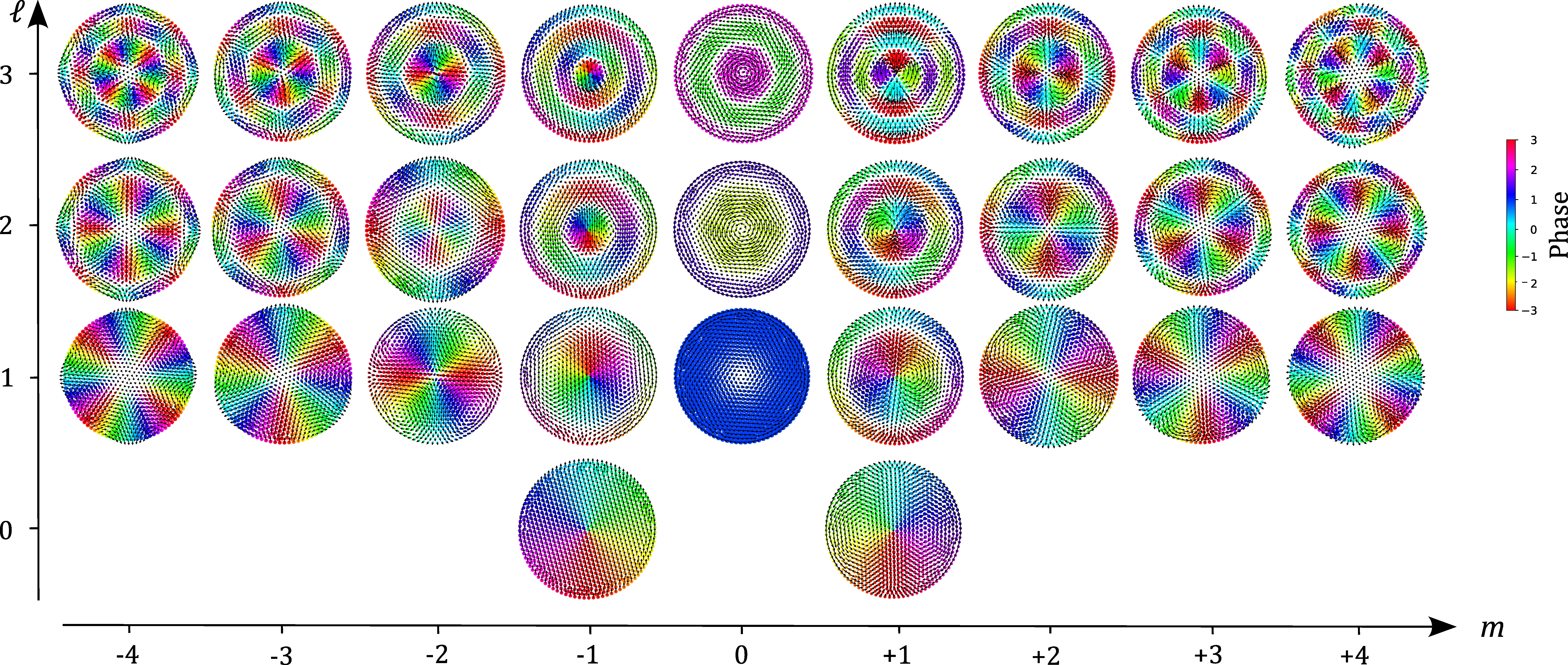}}
    \caption{\label{fig:all-modes-tree}Spatial mode profiles of $m_r$ sorted by their number of radial nodes $\ell$ and azimuthal nodes $m$. The phase and the magnitude are encoded such as in Fig.~5
.}
\end{figure*}

\let\clearpage\relax


%

\end{document}